\documentclass[acus]{JAC2001}
\usepackage{graphicx}
\begin{document}

\thispagestyle{empty}

\onecolumn

\begin{flushright}
{\large
SLAC--PUB--9254\\
June 2002\\}
\end{flushright}

\vspace{.8cm}

\begin{center}

{\LARGE\bf
NLC Luminosity as a Function of Beam Parameters\footnote
{\normalsize{Work supported by
Department of Energy contract  DE--AC03--76SF00515.}}}

\vspace{1cm}

\large{
Y.~Nosochkov, P.~Raimondi, T.O.~Raubenheimer, A.~Seryi \\
Stanford Linear Accelerator Center, Stanford University,
Stanford, CA 94309}

\end{center}

\vfill

\begin{center}
{\LARGE\bf
Abstract }
\end{center}

\begin{quote}
\large{
Realistic calculation of NLC luminosity has been performed using
particle tracking in DIMAD and beam-beam simulations in GUINEA-PIG code for
various values of beam emittance, energy and beta functions at the
Interaction Point (IP).  Results of the simulations are compared with
analytic luminosity calculations.  The optimum range of IP beta functions
for high luminosity was identified.
}
\end{quote}

\vfill

\begin{center}
\large{
{\it Presented at the 8th European Particle Accelerator Conference 
(EPAC 2002)\\
Paris, France, June 3--7, 2002} \\
}
\end{center}

\newpage

\pagenumbering{arabic}
\pagestyle{plain}

\twocolumn

\title{
NLC LUMINOSITY AS A FUNCTION OF BEAM PARAMETERS~\thanks
{Work supported by Department of Energy contract 
DE--AC03--76SF00515.}\vspace{-2mm}}

\author{
Y.~Nosochkov, P.~Raimondi, T.O.~Raubenheimer, A.~Seryi,
SLAC, CA 94309, USA\vspace{0mm}}

\maketitle

\begin{abstract} 

Realistic calculation of NLC luminosity has been performed using
particle tracking in DIMAD and beam-beam simulations in GUINEA-PIG code for
various values of beam emittance, energy and beta functions at the
Interaction Point (IP).  Results of the simulations are compared with
analytic luminosity calculations.  The optimum range of IP beta functions
for high luminosity was identified.

\end{abstract}

\vspace{-1mm}
\section{INTRODUCTION}

Maximizing luminosity is an important part of the NLC design optimization.
Luminosity for the head-on symmetric gaussian beam collisions is given
by~\cite{chen}
\vspace{-2mm}
\begin{equation}
L=\frac{nN^{2}fH}{4\pi\sigma_x^*\sigma_y^*}\,,
\label{eq:lumi}
\vspace{-2mm}
\end{equation}
where $n$, $N$, $f$, $\sigma^*$ and $H$ are, respectively, the number of
bunches per train, number of particles per bunch, repetition rate, {\it
rms} beam size at IP, and luminosity enhancement factor due to pinch and
hourglass effects~\cite{chen}.  For zero IP dispersion, the beam size is
$\sigma^*\!=\!\sqrt{\epsilon^*\beta^*}$, where $\epsilon^*$ and $\beta^*$
are the beam emittance and beta function at IP.

For optimization of luminosity it is desirable to analyze it as a function
of beam parameters such as $\beta^*$ and incoming beam emittance
$\epsilon_0$.  The use of analytical formula (\ref{eq:lumi}) requires
accurate estimate of the final beam emittance $\epsilon^*$ and enhancement
factor $H$.  However, analytical calculation of emittance dilution is
complicated by the combined effect of high order chromaticity, non-linear
fields in the Final Focus (FF) system and synchrotron radiation.
Similarly, accuracy of the empirical formula for luminosity enhancement
factor $H$ \cite{chen} has limitations as well.

For a more accurate and realistic computation of luminosity, we performed
numerical tracking and beam-beam simulations using DIMAD~\cite{dimad} and
GUINEA-PIG~\cite{g-pig} codes.  An automatic routine based on
FFADA~\cite{ffada} was used to generate and track particles in DIMAD and
then use the resultant distribution at IP for beam-beam simulation in
GUINEA-PIG.  Below we present results of these simulations, compare them
with analytical calculations, and examine the optimum range for IP beta
functions.

\vspace{-1mm}
\section{SIMULATIONS}

\begin{figure}[t]
\centering
\includegraphics*[width=60mm, angle=-90]{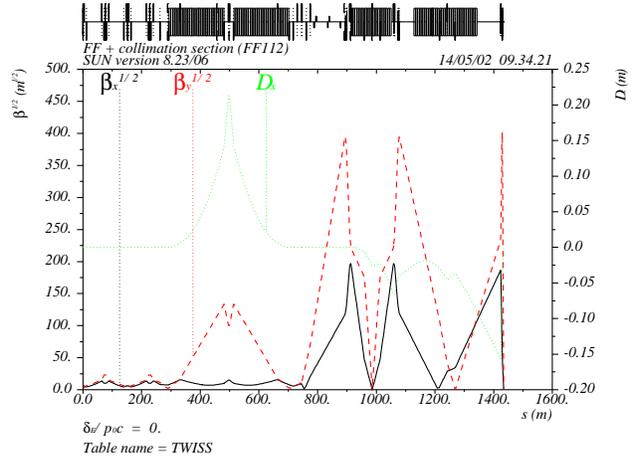}
\vspace{-2mm}
\caption{$\sqrt{\beta_{x,y}}$ functions and dispersion $D_x$
in the collimation and FF sections (IP is on the right).}
\label{fig:optics}
\vspace{-3mm}
\end{figure}

The typical simulation in DIMAD and GUINEA-PIG included 20,000 particles
per beam with the initial gaussian distribution in phase space, except for
the energy spread where an appropriate double-horned distribution was used.
Particles were tracked through the last 1433.8~meters of the NLC beamline
which includes the final collimation section and the FF system as shown in
Fig.~\ref{fig:optics}~\cite{ffs}.  The ideal lattice without magnet errors
was used, and the incoming beam distribution was matched to the initial
machine phase ellipse.  The following ``nominal'' set of the NLC parameters
for 250~GeV beams was used in the calculations:  $n\!=\!192$,
$N\!=\!7.5\!\cdot\!10^9$, $f\!=\!120$~Hz,
$\gamma\epsilon_{x0}/\gamma\epsilon_{y0}\!=\!360/4$~[$10^{-8}$~m],
$\beta_x^*/\beta_y^*\!=\!8/0.11$~mm, and $\sigma_z\!=\!0.11$~mm, where
$\gamma\epsilon_0$ and $\sigma_z$ are the normalized emittance and {\it
rms} bunch length of the incoming beam, respectively.  Emittance growth due
to synchrotron radiation in bends and quadrupoles was included in the DIMAD
tracking, and head-on collisions were assumed for luminosity calculation.

The NLC FF optics includes a non-linear correction system of sextupoles,
octupoles and decapoles for compensation of high order chromatic and
geometric effects generated in the FF and collimation section~\cite{ffs}.
Normally, this system is optimized for the nominal set of parameters.
Variation of emittance, energy and beta functions changes the beam size and
synchrotron radiation effects in the FF.  As a result, an additional tuning
of the non-linear correctors may be needed to maintain maximum luminosity.
However, a complete optimization of this system is somewhat time consuming.
For this reason, a different technique was employed in the simulations to
optimize this compensation.  In this method, a scaling factor was used to
vary all bending angles in the FF dipoles and another factor to scale
dipoles in the collimation section.  Relationship between the two factors
was fixed to keep the IP horizontal position unchanged.  Horizontal
dispersion was linearly changed by the scaling, but remained separately
closed in the collimation and FF sections.  Strengths of the non-linear
correctors were scaled the opposite way to keep chromatic correction near
optimum.  This variation of corrector strengths allows optimization of the
FF geometric aberrations.  The peak luminosity found by optimizing the bend
scale factor is denoted by $L_s$.

\vspace{-1mm}
\subsection{Variation of $\beta^*$ and Emittance}

Luminosity as a function of $\beta^*$ and initial normalized emittance
$\gamma\epsilon_0$ was studied using the NLC FF design for 250~GeV beams
with the nominal parameters listed above.  Two methods for variation of
$\beta^*$ were tested.  In the first method, six matching quadrupoles
located between the collimation section and FF were adjusted to provide a
local optics change for a desired $\beta^*$.  This adjustment changes beta
functions only in the FF system.  As a result, the optimal relationship
between the collimation and FF sections is changed which may reduce the
effect of the non-linear compensation.  In the second method, the FF
quadrupoles were not changed, but the initial betatron functions at the
beginning of collimation section were adjusted to provide a desired
$\beta^*$.  This technique requires that the optics match is done upstream
of the collimation section.  In this method, beta functions change
simultaneously in the collimation and FF sections and, therefore,
transformation between the two sections is preserved which may be a better
option for keeping the non-linear compensation at optimum.  Another
advantage of this approach is that the collimator settings would not need
to be changed.  Simulations showed that luminosity as a function of
$\beta_x^*$ is about the same in both methods, but the second method
delivered up to 7\% higher luminosity at very low $\beta_y^*$.  In the
following sections, only the results of the second method are presented.

\begin{figure}[t]
\centering
\includegraphics*[width=70mm]{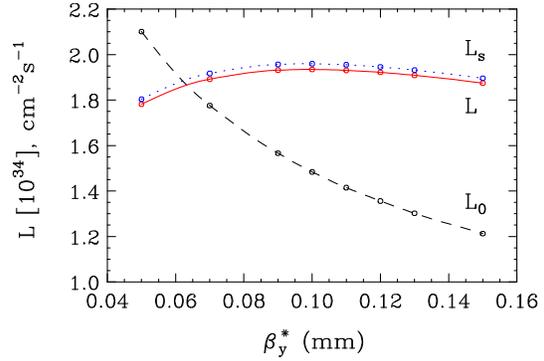}
\vspace{-2mm}
\caption{Luminosity versus $\beta_y^*$.}
\label{fig:lum-by}
\vspace{-0mm}
\end{figure}

\begin{figure}[t]
\centering
\includegraphics*[width=70mm]{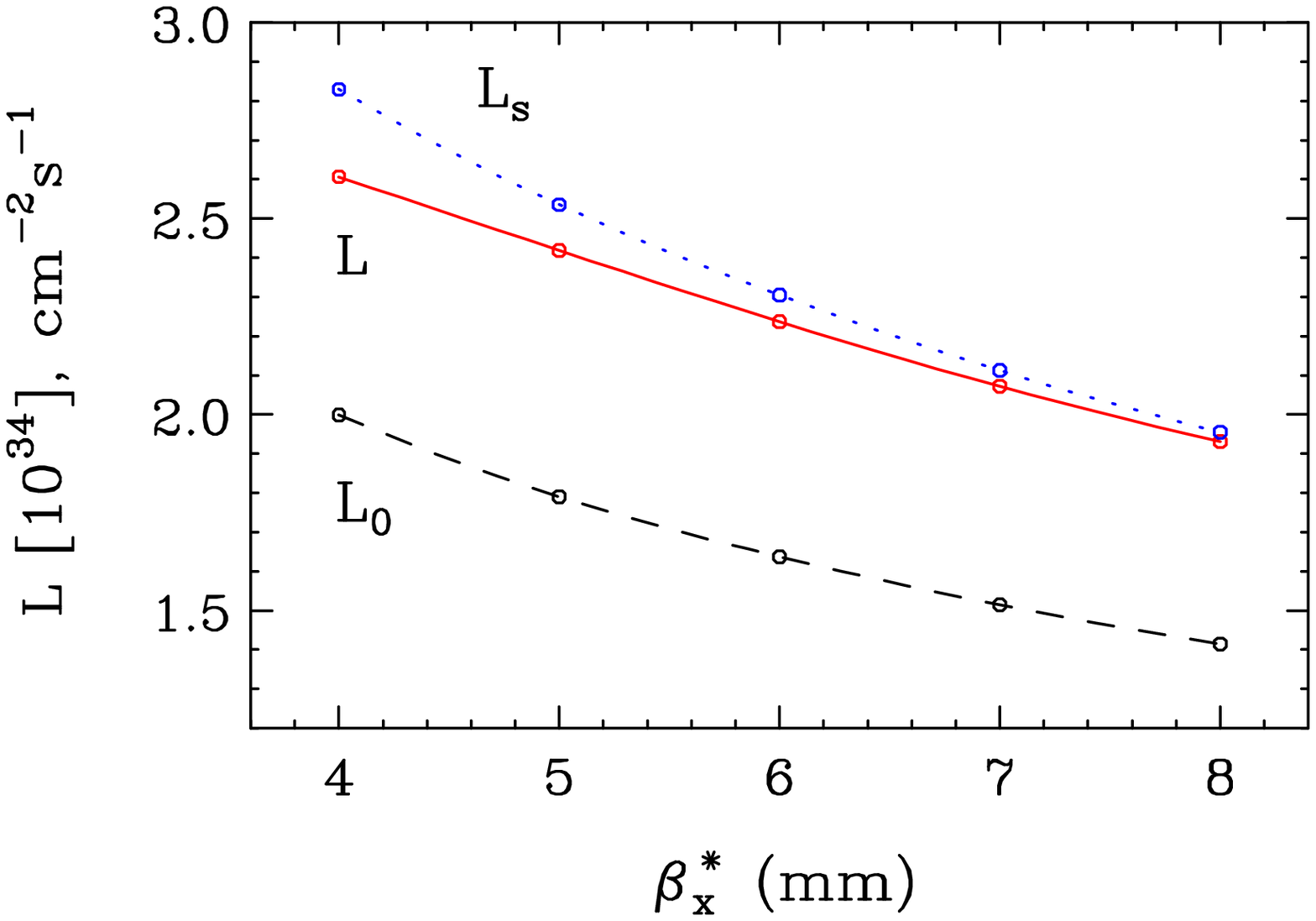}
\vspace{-2mm}
\caption{Luminosity versus $\beta_x^*$.}
\label{fig:lum-bx}
\vspace{-3mm}
\end{figure}

Luminosity versus $\beta_y^*$ and $\beta_x^*$ for 250~GeV beams
is shown in Fig.~\ref{fig:lum-by},~\ref{fig:lum-bx}, where $L$ is
obtained using tracking and beam-beam simulations, $L_s$ is $L$ enhanced by
optimizing the bend scaling factor, and $L_0$ is the analytical luminosity
without emittance growth and enhancement factor $H$:
\vspace{-2mm}
\begin{equation}
L_{0}=\frac{nN^{2}f}{4\pi\sigma_{x0}^*\sigma_{y0}^*}\,,
\label{eq:lumi0}
\vspace{-2mm}
\end{equation}
where $\sigma_0^*\!=\!\sqrt{\epsilon_0^*\beta^*}$.  Note that
$L_0\!\sim\!1/\sqrt{\beta^*}$ in Fig.~\ref{fig:lum-by}
and~\ref{fig:lum-bx}.  Ratio $L/L_0$ quantifies the combined effect of
luminosity enhancement factor $H$ and emittance growth in
the collimation and FF optics.

\begin{figure}[b]
\centering
\includegraphics*[width=70mm]{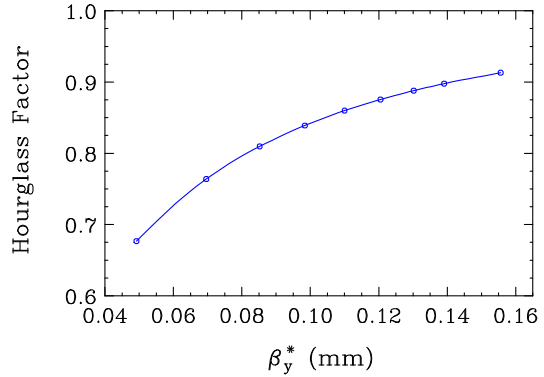}
\vspace{-2mm}
\caption{Hourglass reduction factor for $\sigma_z\!=\!0.11$~mm.}
\label{fig:hglass}
\vspace{-0mm}
\end{figure}

Fig.~\ref{fig:lum-by} shows rather weak dependence of $L$ versus
$\beta_y^*$ with the maximum reached near $\beta_y^*\!=\!0.10$~mm, close to
the nominal value of 0.11~mm.  Luminosity is affected by the beta factor
$1/\sqrt{\beta^*}$, emittance growth (due to synchrotron radiation and high
order aberrations), hourglass reduction and pinch enhancement factors.  At
high values of $\beta_y^*$, luminosity reduction is dominated by the beta
factor, while at low $\beta_y^*$, it is caused by combination of the
hourglass factor, pinch effect and emittance growth which prevail over the
stronger focusing.  The hourglass factor calculated using analytical
formula~\cite{hglass} for $\sigma_z\!=\!0.11$~mm is shown in
Fig.~\ref{fig:hglass} as a function of $\beta_y^*$.  For symmetric flat
beams it only depends on ratio of $\beta_y^*/\sigma_z$.  For example, it
reduces luminosity by 32\% at $\beta_y^*\!=\!0.05$~mm compared to 14\% at
the nominal 0.11~mm.  Emittance growth at low $\beta_y^*$ qualitatively may
be explained by the increased beam size in the FF magnets
($\sim\!1/\sqrt{\beta_y^*}$) which enhances non-linear optical aberrations.
Bend scaling increases luminosity ($L_s$) only by $\sim\!1\%$, but requires
$\sim\!30\%$ stronger bends in the FF.

Luminosity versus $\beta_x^*$ is shown in Fig.~\ref{fig:lum-bx}, where the
nominal $\beta_x^*$ is 8~mm.  In this case, the hourglass factor is roughly
constant.  At low $\beta_x^*$, luminosity is increased by the beta factor
and to less extent by the pinch effect which prevail over reduction due to
emittance growth.  Up to 9\% higher luminosity ($L_s$) can be achieved at
low $\beta_x^*$ by scaling the FF bends by up to +57\%.  The scaling
increases dispersion and reduces non-linear corrector strengths required
for chromatic compensation.  As a result, enhancement of geometric
aberrations caused by larger beam size in the FF
($\sim\!1/\sqrt{\beta_x^*}$) is compensated by the reduction of the
sextupole and other corrector strengths.

\begin{figure}[t]
\centering
\includegraphics*[width=70mm]{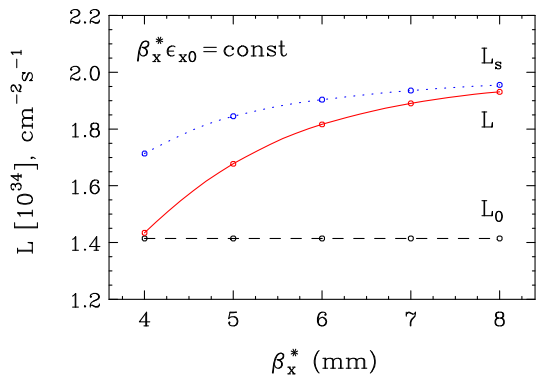}
\vspace{-2mm}
\caption{Luminosity versus $\beta_x^*$ for constant
$\beta_x^*\epsilon_{x0}$.}
\label{fig:lum-bx-emx}
\vspace{-3mm}
\end{figure}

Fig.~\ref{fig:lum-bx-emx} shows luminosity versus $\beta_x^*$, where
$\beta_x^*\epsilon_{x0}$ is kept constant.  The first order beam size at IP
and $L_0$ are constant in this case, but $L$ is significantly reduced at
low $\beta_x^*$.  This is caused by increased emittance growth from
geometric aberrations which are enhanced by larger beam size in the FF
magnets ($\sim\!1/\beta_x^*$).  Similarly to the previous case, luminosity
can be improved by scaling the FF bends by up to +85\% at the lowest
$\beta_x^*$.

\vspace{-1mm}
\subsection{Variation of Beam Energy}

\begin{figure}[t]
\centering
\includegraphics*[width=70mm]{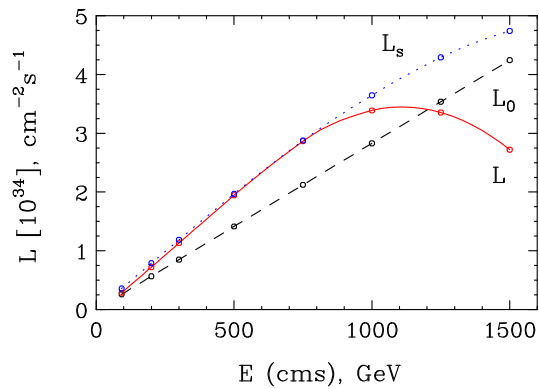}
\vspace{-2mm}
\caption{Luminosity versus beam energy ({\it cms}).}
\label{fig:lum-ener}
\vspace{-0mm}
\end{figure}

Luminosity as a function of energy ({\it cms}) is shown in
Fig.~\ref{fig:lum-ener}, where nominal parameters for 250~GeV beams are
used for all energy variation.  A significant reduction of $L$ at high
energy is explained by the effects of synchrotron radiation in the FF bends
and quadrupoles.  As the energy $E$ increases, the beam size decreases as
$1/\sqrt{E}$ which reduces high order geometric aberrations.  But energy
spread created by synchrotron radiation in bends increases with energy and
enhances chromatic aberrations.  This results in the luminosity loss if the
non-linear correction is not reoptimized.  Using the bend scaling method,
compensation can be improved and most of the luminosity recovered.  For
peak luminosity $L_s$ at high energy, the bending angles were scaled down
which reduced dispersion and energy spread from synchrotron radiation.  The
corrector strengths were correspondingly scaled up, but it was acceptable
since geometric aberrations were reduced due to the smaller beam size.  The
full range of the scaling factor was from 2.76 at $E\!=\!92$~GeV ({\it
cms}) to 0.42 at 1.5~TeV.  Reoptimization of the final doublet length to
minimize synchrotron radiation effects would keep the luminosity increasing
at energy higher than 1.5~TeV ({\it cms}).

\vspace{-1mm}
\subsection{Comparison with Analytical Calculation}

\begin{figure}[t]
\centering
\includegraphics*[width=70mm]{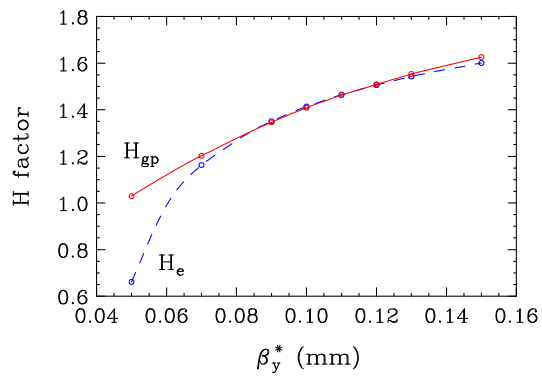}
\vspace{-2mm}
\caption{Luminosity enhancement factor $H$ versus $\beta_y^*$.}
\label{fig:pinch-by}
\vspace{-3mm}
\end{figure}

Accuracy of the empirical formula for luminosity enhancement factor $H$
\cite{chen} was verified by comparison with the GUINEA-PIG simulations.
The $H$ factor includes the hourglass reduction and pinch enhancement
effects.  Fig.~\ref{fig:pinch-by} shows the GUINEA-PIG factor $H_{gp}$ and
empirical factor $H_e$ versus $\beta_y^*$ corresponding to
Fig.~\ref{fig:lum-by}.  It follows from Ref.~\cite{chen} that $H_e$ was
derived for $\sigma_z/\beta_y^*\!\leq\!0.8$ and its accuracy is
$\sim\!10\%$.  Fig.~\ref{fig:pinch-by} shows an agreement of 1-5\% between
$H_{gp}$ and $H_e$ for $\sigma_z/\beta_y^*\!\leq\!1.5$, but $H_e$
significantly underestimates luminosity at lower $\beta_y^*$.  We conclude
that for practical NLC beam parameters the empirical formula provides
reasonable estimate of luminosity.

\vspace{-1mm}
\section{CONCLUSION}

Tracking and beam-beam simulations in the NLC collimation and FF sections
for 250~GeV beams showed that luminosity is at maximum near
$\beta_y^*\!=\!0.10$~mm and can be further increased by reducing
$\beta_x^*$.  Optimization of the non-linear compensation scheme for
maximum luminosity is required when beam parameters change, especially at
high beam energy.  Accuracy of the empirical formula for luminosity
enhancement factor was found to be reasonably good for the NLC parameters
close to nominal.

\end{document}